\documentclass[12pt]{article}

\usepackage{amsmath}
\usepackage{amsfonts}
\usepackage{amssymb}
\usepackage{amsthm}
\usepackage{stmaryrd} 
\usepackage{braket}
\usepackage{empheq}
\usepackage{graphicx}
\usepackage{subfigure}
\usepackage{hyperref}
\usepackage{footmisc}
\usepackage{appendix}
\usepackage{epsf}
\usepackage{epsfig}
\usepackage{tikz,tikz-3dplot}
\usetikzlibrary{decorations.markings,decorations.pathmorphing}
\usepackage{caption}
\usepackage{pgfplots}
\usepackage[utf8]{inputenc}
\usepackage[english]{babel}
\usepackage[T1]{fontenc}
\usepackage{sectsty}
\usepackage{authblk}
\subsubsectionfont{\normalfont\itshape}
\usepackage{fullpage}

\renewcommand{\bar}{\overline}
\renewcommand{\leq}{\leqslant}
\renewcommand{\geq}{\geqslant}

\newcommand{\bbone}{{\text{\usefont{U}{bbold}{m}{n}\char49}}}

\theoremstyle{plain}
\newtheorem{theorem}{Theorem}[section]
\newtheorem{proposition}[theorem]{Proposition}

\newtheorem{lemma}[theorem]{Lemma}

\theoremstyle{definition}

\newtheorem{remark}[theorem]{Remark}

\title{Why is the universe not frozen by the quantum Zeno effect?}

\author[a b]{\textsc{Antoine Soulas}}
\affil[a]{Faculty of Physics, University of Vienna \\ Boltzmanngasse 5, 1090 Vienna (Austria); antoine.soulas@univie.ac.at}
\affil[b]{IQOQI Vienna, Austrian Academy of Sciences, Boltzmanngasse 3, 1090 Vienna (Austria)}
\date{}

\date{}

\begin{document}
\maketitle

\abstract{We build a discrete model that simulates the ubiquitous competition between the free internal evolution of a two-level system and the decoherence induced by the interaction with its surrounding environment. It is aimed at being as universal as possible, so that no specific Hamiltonian is assumed. This leads to an analytic criterion, depending on the level of short time decoherence, allowing to determine whether the system will freeze due to the Zeno effect. We check this criterion on several classes of functions which correspond to different physical situations. In the most generic case, the free evolution wins over decoherence, thereby explaining why the universe is indeed not frozen. We finally make a quantitative comparison with the continuous model of Presilla, Onofrio and Tambini, based on a Lindblad's master equation, a find good agreement at least in the low coupling regime.} 


\section{Introduction} \label{intro}

The Zeno effect typically occurs when a quantum system is repeatedly measured: if the time interval between two successive measurements tends to 0, the evolution of the system gets frozen. The main reason is that, in quantum mechanics, the general short time evolution is quadratic, \textit{i.e.}: 
\[ \lvert \braket{\Psi \vert \Psi(t)} \rvert^2 = \lvert \braket{\Psi \vert e^{-i\hat{H}t} \vert \Psi} \rvert^2 = 1 - V t^2 + O\left(t^4\right), \] 
where $V \equiv \mathrm{Var}_{\ket{\Psi}}(\hat{H}) = \braket{\Psi \vert \hat{H}^2 \vert \Psi} - \braket{\Psi \vert \hat{H}  \vert \Psi}^2$ (we take $\hbar = 1$). Hence, if $n$ projective measurements along $\ket{\Psi}$ are performed during a fixed time interval $T$, the probability $p_n$ that all the measurements gave the outcome $\ket{\Psi}$ is, at leading order:
\[ p_n \simeq \left(1 - V \left( \frac{T}{n} \right)^2 \right)^n \underset{n \rightarrow +\infty} \longrightarrow 1. \]
Note that to obtain this limit, one has to neglect the higher order terms, as is usually done in the standard presentations of the Zeno effect \cite[\S 3]{pascazio2014all} \cite[\S 3.3.1.1]{joos1996decoherence}. Rigorously speaking, this is an additional assumption, because when developing the expression $\left(1 - V \left( \frac{T}{n} \right)^2 + O\left( \frac{1}{n^4} \right) \right)^n$, the number of $O\left( \frac{1}{n^4} \right)$ actually depends on $n$.

In the spirit of the theory of decoherence, one might wish to get rid of the ill-defined notion of (ideal) projective measurement. Since a measurement is nothing but a particular case of interaction with an environment that entails strong decoherence of the system in the measured basis, it is tempting to ask what level of decoherence is required to freeze a system. For example, a particle in a gas is continuously monitored ($\sim$ measured) by its neighbors, yet the gas manifestly has an internal evolution... and so does the universe in general. It is not obvious \textit{a priori} whether quantum mechanics actually predicts that the universe is not frozen. 

This question has already been addressed by examining the continuous dynamics of the pair system - environment for relatively generic Hamiltonian \cite{joos1984continuous}. The Zeno limit is recovered for strong interaction, and Fermi's golden rule is recovered in the limit of small interaction. ‘The model shows that the coupling to the environment leads to constant transition rates which are unaffected by the measurement if the coupling is "coarse enough" to discriminate only between macroscopic properties. This may in turn be used to define what qualifies a property as macroscopic: it must be robust against monitoring by the environment’ \cite[\S 3.3.2.1]{joos1996decoherence}. Similarly, the master equation for the motion of a mass point under continuous measurement indicates that the latter is not slowed down because the Ehrenfest theorems are still valid. ‘This may be understood as a consequence of the fact that, for a continuous degree of freedom, any measurement with finite resolution necessarily is too coarse to invoke the Zeno effect’ \cite[\S 3.3.1.1]{joos1996decoherence}.

In addition, Presilla, Onofrio and Tambini have studied how the continuous monitoring of a two-level system affects its time evolution and its coherence based on a Lindblad's master equation \cite[\S IV.]{presilla1996measurement}. They have in particular confronted their results to that of a historical experimental test of the Zeno effect \cite{itano1990quantum}, and were able to obtain a better fit of the data than the naïve approach (assuming perfect von Neumann collapses) by taking into account the finite duration of the measurements. We will come back to this approach in Section \ref{comparison} to make a quantitative comparison between the latter's results and ours.

Another interesting model is that of \cite[\S8.3 and \S8.4]{fonda1978decay}. It may at first sight seem puzzling that an unstable nucleus continuously measured by a Geiger counter can actually decay. Indeed, if the measurement is treated as an ideal projective one, the nucleus should continuously be projected onto a non-decayed state. But as soon as the decoherence process is not supposed immediate anymore (even as short as $10^{-16}$s, see equation (8.45) in \cite{fonda1978decay}), the deviation from the expected exponential decay is shown to be negligible.

Although these models are already convincing, our aim is to give a new contribution to this topic of understanding why the vast majority of physics is not affected by the quantum Zeno effect, the latter being detectable only in some very specific experimental setups. Our model also formalizes the competition between free evolution (no information leaking to the rest of the world) and decoherence (interaction with the environment), but differs from the previous ones in two respects: its mathematical structure is discrete and it does not assume anything about the form of the Hamiltonian, so as to be as universal as possible. The use of a discrete framework is consistent with the approach adopted in a lot of mathematical studies on the quantum Zeno effect (see \cite{mobus2023optimal} and references therein).

\section{The model: free evolution \textit{vs.} decoherence} \label{model_zeno}

Having in mind the fact that continuous degrees of freedom are less prone to the Zeno effect (recall the previous quote from \cite{joos1996decoherence}), in order to explain why the universe is not frozen, it may suffice to check it on a two-level system. Our system of interest will therefore be a qbit, initially in the state $\ket{0}$ and monitored by an environment producing partial decoherence in the basis $(\ket{0} , \ket{1})$. We consider a fixed time interval $T$, divided into $n$ phases of length $\delta = \frac{T}{n}$ dominated by the free evolution. This evolution takes the general form:
\[ \left\{ \begin{array}{ll}
 U_\delta \ket{0} = c_{=}^{0}(\delta) \ket{0} + c_{\neq}^{1}(\delta) \ket{1}  \\ \\
U_\delta \ket{1} = c_{\neq}^{0}(\delta) \ket{0} + c_{=}^{1}(\delta) \ket{1},
 \end{array} \right. \] 
where the coefficients satisfy $\lvert c_{=}^{0}(\delta) \rvert^2 =  \lvert c_{=}^{1}(\delta) \rvert^2 = 1 - V\delta^2 +  O\left(\delta^4\right)$ and $\lvert c_{\neq}^{0}(\delta) \rvert^2 =  \lvert c_{\neq}^{1}(\delta) \rvert^2 = V\delta^2 +  O\left(\delta^4\right)$ (in the sequel, we will drop the argument $\delta$ whenever the context is clear). As recalled in the introduction, to stick to the standard derivations of the Zeno effect, we need to neglect all the higher order terms, so that we actually suppose $\lvert c_{=}^{0}(\delta) \rvert^2 =  \lvert c_{=}^{1}(\delta) \rvert^2 = 1 - V\delta^2$ and $\lvert c_{\neq}^{0}(\delta) \rvert^2 =  \lvert c_{\neq}^{1}(\delta) \rvert^2 = V\delta^2 $.

\begin{figure}[h] 
\centering
\includegraphics[scale=0.25]{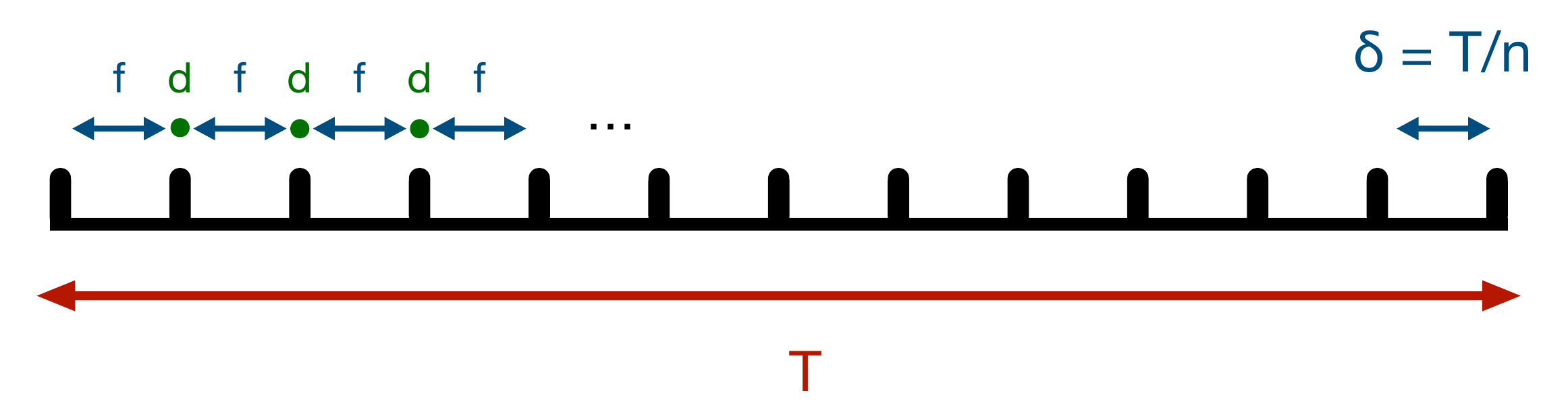}
\caption{\small{Alternating steps of free evolution (f) and decoherence (d)}}
\label{fig1}
\end{figure}

After the i$^{\text{th}}$ phase of free evolution, the system meets some neighboring environment $\mathcal{E}^i$, initially in the state $\ket{\mathcal{E}^i_\text{init} }$, and gets immediately entangled according to:
\[ \left\{ \begin{array}{ll}
 \ket{0} \longrightarrow \ket{0} \ket{\mathcal{E}^i_{0}} \\ \\
\ket{1} \longrightarrow \ket{1} \ket{\mathcal{E}^i_{1}},
 \end{array} \right. \] 
where $\lvert \braket{\mathcal{E}^i_{0} \vert \mathcal{E}^i_{1}} \rvert \equiv \eta^i$ quantifies the level of decoherence induced by $\mathcal{E}^i$, \textit{i.e.} how well the environment has recorded the system's state ($\eta^i = 1$ means no decoherence, $\eta^i = 0$ perfect decoherence). See Figure \ref{fig1}.

From now on, we suppose that $\eta^i \equiv \eta$ does not depend on $i$ (taken as a mean level of decoherence), which amounts to assuming that the strength of the interaction is more or less constant over time. Finally, we also suppose that each environment $\mathcal{E}^i$ is distinct from the others and non-entangled at the time it encounters the system.

Recalling that $T=n\delta$, the relevant quantity to compute is the probability $p_n$ that, at the end of the time interval $T$, the system is still found in its initial state $\ket{0}$ and that all the successive environments have recorded $0$.

\begin{proposition}
Neglecting all the higher order terms, we can write:
\begin{align*}
p_n \simeq 1 - 2 \Big[ \frac{n}{2} + (n-1)\eta + (n-2)\eta^2 + \dots + \eta^{n-1} \Big] V \delta^2.
\end{align*}
\end{proposition}

\begin{proof}
The cases $n=1$ or 2 are easy to treat. Indeed, the successive iterations go as follows ($f$ stands for the free evolution and $d$ for the decoherence step):

\begin{align*}
\ket{0} \equiv \ket{\Psi_0} & \overset{f} \leadsto c_{=}^{0} \ket{0} + c_{\neq}^{1} \ket{1} \\
& \overset{d} \leadsto  c_{=}^{0} \ket{0} \ket{\mathcal{E}^1_{0}} + c_{\neq}^{1} \ket{1} \ket{\mathcal{E}^1_{1}} \equiv \ket{\Psi_1} \\ \\
& \overset{f} \leadsto c_{=}^{0} \Big[c_{=}^{0} \ket{0} + c_{\neq}^{1} \ket{1} \Big] \ket{\mathcal{E}^1_{0}} + c_{\neq}^{1} \Big[c_{\neq}^{0} \ket{0} + c_{=}^{1} \ket{1} \Big] \ket{\mathcal{E}^1_{1}} \\
& \quad \quad = \ket{0} \Big[ c_{=}^{0}c_{=}^{0} \ket{\mathcal{E}^1_{0}} + c_{\neq}^{0} c_{\neq}^{1} \ket{\mathcal{E}^1_{1}} \Big] +  \ket{1} \Big[ c_{\neq}^{1} c_{=}^{0} \ket{\mathcal{E}^1_{0}} + c_{=}^{1} c_{\neq}^{1} \ket{\mathcal{E}^1_{1}} \Big] \\
& \overset{d} \leadsto \ket{0} \Big[ c_{=}^{0}c_{=}^{0} \ket{\mathcal{E}^1_{0}} + c_{\neq}^{0} c_{\neq}^{1} \ket{\mathcal{E}^1_{1}} \Big] \ket{\mathcal{E}^2_{0}} +  \ket{1} \Big[ c_{\neq}^{1} c_{=}^{0} \ket{\mathcal{E}^1_{0}} + c_{=}^{1} c_{\neq}^{1} \ket{\mathcal{E}^1_{1}} \Big] \ket{\mathcal{E}^2_{1}} \equiv \ket{\Psi_2}.
\end{align*}
The $(\ket{\Psi_n})_{n \in \mathbb{N}}$ seem to live in different Hilbert spaces only because we omit to write all the environments $(\mathcal{E}^i)_{i\geq n+1}$ with which the system is not entangled yet. Consequently, neglecting all the higher order terms yields:
\begin{align*}
\bullet \; p_1 &= \left\lvert \braket{ 0 \mathcal{E}^1_{0}  \vert \Psi_1} \right\rvert^2 = \lvert c_{=}^{0} \rvert^2 = 1 - V \delta^2 \\
\bullet \; p_2 &= \left\lvert \braket{ 0 \mathcal{E}^1_{0} \mathcal{E}^2_{0}  \vert \Psi_2} \right\rvert^2 \\
&= \left\lvert c_{=}^{0 2} +  c_{\neq}^{0} c_{\neq}^{1} \braket{\mathcal{E}^1_{0} \vert \mathcal{E}^1_{1}} \right\rvert^2 \\
&= (1-V\delta^2)^2 + \eta^2 (V\delta^2)^2 + 2 \Re\Big(\bar{c_{=}^{0}}^2 c_{\neq}^{0} c_{\neq}^{1} \braket{\mathcal{E}^1_{0} \vert \mathcal{E}^1_{1}} \Big) \\
&\simeq 1 - 2 (1+\eta)V\delta^2.
\end{align*}
The last step is not obvious and comes from the following argument. \textit{A priori}, the quantity $\Re\Big(\bar{c_{=}^{0}}^2 c_{\neq}^{0} c_{\neq}^{1} \braket{\mathcal{E}^1_{0} \vert \mathcal{E}^1_{1}} \Big)$ lies in $[-\eta V \delta^2 , \eta V \delta^2]$ up to a $O\left(\delta^4\right)$, but the coefficients of the matrix $ U_\delta$ are not unrelated. Using the general parametrization of a $2\times2$ unitary matrix, $U_\delta = \begin{pmatrix}  c_{=}^{0} &  c_{\neq}^{0} \\  c_{\neq}^{1} &  c_{=}^{1} \end{pmatrix}$ can be written $\begin{pmatrix} a & b \\ -e^{i\varphi} \bar{b} & e^{i\varphi} \bar{a} \end{pmatrix}$. Moreover, for small $\delta$ (this approximation may be rough for the case $n=2$ but gets better as $n$ increases),  $U_{\delta} \rightarrow \bbone$ hence $\det(U_{\delta}) = e^{i\varphi} \rightarrow 1$ and $\bar{a} \rightarrow 1$. We also expect $\braket{\mathcal{E}^1_{0} \vert \mathcal{E}^1_{1}}$ to be close to the real number 1 (infinitesimal decoherence). Combining all this, $\bar{c_{=}^{0}}^2 c_{\neq}^{0} c_{\neq}^{1} \braket{\mathcal{E}^1_{0} \vert \mathcal{E}^1_{1}} = - e^{i \varphi} \bar{a}^2 \lvert b \rvert^2 \braket{\mathcal{E}^1_{0} \vert \mathcal{E}^1_{1}}$ is close to be a real negative number, therefore its real part is approximately the opposite of its modulus. \\

In general, $p_n = \left\lvert \braket{ 0 \mathcal{E}^1_{0} \dots \mathcal{E}^n_{0}  \vert \Psi_n} \right\rvert^2$ is the square modulus of a sum of terms of the form 
\[ z_\alpha^b = c_{\alpha_1}^{b_1} \dots c_{\alpha_n}^{b_n} \quad \braket{\mathcal{E}^1_{0} \vert \mathcal{E}^1_{b_1}} \dots \braket{\mathcal{E}^n_{0} \vert \mathcal{E}^n_{b_n}}, \]
where $\alpha = \alpha_1 \dots \alpha_n$ and $b = b_1 \dots b_n$ are words on the alphabets $\{= , \neq\}$ and $\{0,1\}$ respectively. The word $b$ is entirely deduced from $\alpha_1 \dots \alpha_i$ according to: 
\begin{align*} 
b_0 = 0 \quad ; \quad b_{i} = \left\{ \begin{array}{ll}
 b_{i-1} \quad &\text{ if } \alpha_i \text{ is  = (state preserved)} \\
 b_{i-1} + 1 \mod 2  \quad &\text{ if } \alpha_i \text{ is $\neq$ (state flipped),}
 \end{array} \right. 
 \end{align*} 
with the additional requirement that $b_n=0$ (system finally measured in state $\ket{0}$), so that $\alpha$ actually contains an even number of $\neq$. Note that only the indices $i$ such that $b_i=1$ contribute non-trivially in the product of brackets, since $\braket{\mathcal{E}^i_{0} \vert \mathcal{E}^i_{0}}=1$.

We now use the fact that $\left\lvert \sum_k z_k \right\rvert^2 = \sum_k \left\lvert z_k \right\rvert^2 + \sum_{k<l} 2\Re(\bar{z_k} z_l)$ for all complex numbers $(z_k)_k$. In our case, the leading term is clearly $\lvert z_{=\ldots=}^{0\dots 0} \rvert^2 = \lvert {c_{=}^0}^n \rvert^2 = (1-V\delta^2)^n \simeq 1- nV\delta^2$, while all the other square moduli are of order $\delta^4$ or less because they contain at least two factors $\lvert c_{\neq}^{b_i}\rvert^2 = V\delta^2$. Furthermore, repeating the above argument, the real parts can be approximately replaced by their opposite moduli (and this approximation is better as $n$ gets larger). Therefore, only the cross-products of the form $2\Re(\bar{{c_{=}^{0}}^n} \times z_\alpha^b)$, where $\alpha$ contains exactly two $\neq$, contribute at order $\delta^2$. The power of $\eta$ that appears in this cross-product (\textit{i.e.} the number of non-trivial brackets $\braket{\mathcal{E}^i_{0} \vert \mathcal{E}^i_{1}}$) is the number of indices $i$ such that $b_i=1$, that is the number of steps elapsed between the two $\neq$. For instance, if the two $\neq$ happen at the $i^{\text{th}}$ and $j^{\text{th}}$ step, the contribution is:
\begin{align*}
&2 \Re(\bar{{c_{=}^{0}}^n} \times {c_{=}^0} ^{i-1} c_{\neq}^{1} {c_{=}^{1}}^{j-i-1}c_{\neq}^{0} {c_{=}^{0}}^{n-j-1} \braket{\mathcal{E}^i_{0} \vert \mathcal{E}^i_{1}} \dots \braket{\mathcal{E}^{j-1}_{0} \vert \mathcal{E}^{j-1}_{1}}) \\
\simeq &2 \lvert c_{\neq}^{1} c_{\neq}^{0}  \braket{\mathcal{E}^i_{0} \vert \mathcal{E}^i_{1}} \dots \braket{\mathcal{E}^{j-1}_{0} \vert \mathcal{E}^{j-1}_{1}} \rvert \\
\simeq &2 \eta^{j-i} V \delta^2.
\end{align*}
There are obviously $n-k$ words $\alpha$ with exactly two $\neq$ separated by $k$ steps, corresponding to the $n-k$ possible choices for $i$, whose contribution is $2 \eta^k V \delta^2$. Finally, the general expression for $p_n$ when neglecting all the higher order terms is:
\[ p_n = \left\lvert \braket{ 0 \mathcal{E}^1_{0} \dots \mathcal{E}^n_{0}  \vert \Psi_n} \right\rvert^2 \simeq 1 - 2 \Big[ \frac{n}{2} + (n-1)\eta + (n-2)\eta^2 + \dots + \eta^{n-1} \Big] V \delta^2. \]
\end{proof}

We can check the consistency of this result on two particular cases:
\begin{itemize}
\item \underline{if $\eta=1$}, no decoherence occurs, so we recover the free evolution case during a time interval $n\delta$ instead of $\delta$, \textit{i.e.} $\mathbb{P} _n = 1 - V (n\delta)^2$;
\item \underline{if $\eta=0$}, a perfect decoherence means that the environment acts as an ideal measuring device, so we recover the Zeno case recalled in the introduction, that is $p_n = 1 - n V \delta^2 \simeq (1 - V \delta^2 )^n$.
\end{itemize}

Now, a Zeno effect will freeze the system in the limit of large $n$ if and only if $p_n \underset{n \rightarrow +\infty} \longrightarrow 1$, that is (using $\delta = \frac{T}{n}$) if $\frac{(n-1)\eta + (n-2)\eta^2 + \dots + \eta^{n-1}}{n^2} \underset{n \rightarrow +\infty} \longrightarrow 0$. After some algebra, this expression can be simplified and leads to the following criterion:
\[ \boxed{\text{Zeno effect} \Longleftrightarrow \frac{n\eta(1-\eta) + \eta (\eta^n-1)}{n^2 (1-\eta)^2} \underset{n \rightarrow +\infty} \longrightarrow 0}  \]

We immediately note that if $\eta \in [0,1[$ is a constant independent of $n$, the criterion is satisfied. This is natural because, as the duration of each free evolution phase goes to 0, a constant (even weak) decoherence is applied infinitely many times, so the system freezes.

From now on, we will suppose that the level of decoherence depends on $n$, with $\eta_n \underset{n \rightarrow +\infty} \longrightarrow 1$. A global factor $\eta$ can thus be dropped in the above criterion. Our task in the following sections will be (i) to check the criterion on some common classes of functions $\eta_n$ (section \S\ref{analysis}) (ii) to estimate the level of decoherence really encountered in physical situations (section \S\ref{physical}). \\

\begin{remark}
How finely should the time interval be divided so that the quadratic approximation be valid? Let's forget for a moment that our system is finite dimensional and consider the Hamiltonian of a free particle $\frac{\hat{P}^2}{2m}$, starting from the initial state $\ket{\Psi}(p)=\sqrt{\frac{\sigma}{\sqrt{\pi}\hbar}} e^{-\frac{p^2 \sigma^2}{2 \hbar^2}}$ centred around $x=0$ and $p=0$, and compute:
\begin{align*}
\mathrm{Var}_{\ket{\Psi}}(\hat{H}) &= \frac{1}{4m^2} \left[ \braket{\Psi \vert \hat{P}^4 \vert \Psi} - \braket{\Psi \vert \hat{P}^2  \vert \Psi}^2 \right] \\
&= \frac{1}{4m^2} \left[ \int_{-\infty}^{+\infty} p^4 \frac{\sigma}{\sqrt{\pi} \hbar} e^{-\frac{p^2 \sigma^2}{\hbar^2}} \mathrm{d}p -  \left(\int_{-\infty}^{+\infty} p^2 \frac{\sigma}{\sqrt{\pi} \hbar} e^{-\frac{p^2 \sigma^2}{\hbar^2}} \mathrm{d}p \right)^2 \right] \\
&= \frac{\hbar^4}{8m^2 \sigma^4}
\end{align*}
Hence the quadratic approximation is valid for times shorter than $t_c = \frac{\hbar}{\sqrt{\mathrm{Var}_{\ket{\Psi}}(\hat{H})}} = \frac{2\sqrt{2}m\sigma^2}{\hbar}$. Taking for instance $m = 10^{-26}$kg and $\sigma = 10^{-10}$m, we get $t_c = 4 . 10^{-13}$s. This is way shorter than the mean free time of a particle in a gas in standard conditions, which is of order $10^{-10}$s. So it seems at first sight that the decoherence steps could in practice be too separated in time for the quadratic approximation to be valid all along the free evolution step. However, decoherence doesn't need any actual interaction to take place (a ‘null measurement’ is still a measurement \cite{kwiat1995interaction}). The fact that all the other surrounding particles do \textit{not} interact with the particle of interest is still a gain of information for the environment, which suffices to suppress coherence with other possible histories in which they \textit{would} have interacted. In this case, information is continually leaking to the environment, so it seems legitimate to divide the time interval $T$ as finely as desired so that the quadratic approximation become valid, and the resulting behaviour is then determined by the intensity of infinitesimal decoherence only. The philosophy of this argument is not specific to the infinite dimensional case, and may be applied to our two-level system. It relies, however, on the already mentionned assumption that the strength of the interaction is more or less constant over time. This will be discussed in Section \S\ref{discussion}.
\end{remark}

\section{Analytic study of the criterion} \label{analysis}
Whenever $(n(1-\eta_n))_{n\in\mathbb{N}}$ admits a limit in $\bar{\mathbb{R}}_+ \equiv \mathbb{R}_+ \cup \{ +\infty \}$, the following lemma allows to check immediately the criterion of the previous Section.

\begin{lemma}
Suppose $n(1-\eta_n) \underset{n \rightarrow +\infty} \longrightarrow \alpha \in \bar{\mathbb{R}}_+$. Then 

\[ \frac{n(1-\eta_n) + \eta_n^n-1}{n^2 (1-\eta_n)^2}  \underset{n \rightarrow +\infty} \longrightarrow 
\left\{ \begin{array}{lll}
 \frac{1}{2} \quad \text{if $\alpha = 0$} \\ \\
 0 \quad \text{if $\alpha = + \infty$} \\ \\
 \frac{1}{\alpha} + \frac{e^{-\alpha} -1}{\alpha^2} \quad \text{otherwise.}
 \end{array} \right.\]
\end{lemma}

\begin{proof}
Let $u_n \equiv n(1-\eta_n)$. If $u_n \underset{n \rightarrow +\infty} \longrightarrow +\infty$, since $\eta_n^n-1$ is bounded, the result is immediate. If $0 < \alpha < +\infty$, notice that $\eta_n^n = e^{n \ln\left(1-\frac{u_n}{n} \right)}  \underset{n \rightarrow +\infty} \longrightarrow e^{-\alpha}$, and rewrite: 
\[ \frac{n(1-\eta_n) + \eta_n^n-1}{n^2 (1-\eta_n)^2} = \frac{1}{u_n} + \frac{\eta_n^n-1}{u_n^2} \underset{n \rightarrow +\infty} \longrightarrow \frac{1}{\alpha} + \frac{e^{-\alpha} -1}{\alpha^2}.\]
Finally, if $\alpha=0$:
\begin{align*}
\eta_n^n = e^{n \ln\left(1-\frac{u_n}{n} \right)} &= e^{-u_n - u_n^2/2n + O\left(u_n^3 / n^2 \right)} \\
&= 1 - u_n - \frac{u_n^2}{2n} +  O\left( \frac{u_n^3}{n^2} \right) + \frac{1}{2} \left[ -u_n - \frac{u_n^2}{2n} +  O\left( \frac{u_n^3}{n^2} \right)  \right]^2 + O(u_n^3) \\
&= 1 - u_n + \frac{u_n^2}{2} + O\left( \frac{u_n^2}{n} \right).
\end{align*}
Consequently, $\frac{n(1-\eta_n) + \eta_n^n-1}{n^2 (1-\eta_n)^2} = \frac{u_n^2/2 + O\left( u_n^2/n \right) }{u_n^2}  \underset{n \rightarrow +\infty} \longrightarrow \frac{1}{2}.$
\end{proof}

Two natural candidates for the level of short-time decoherence are $\eta_n = 1 - \frac{\alpha}{ n^{\beta}}$ and $\eta_n = 1 - \alpha e^{-\beta n}$ for $\alpha, \beta >0$. These cases can be treated by the lemma, and the different possible situations are summarized in the following table.\\

\begin{tabular}{| c | c | c |}
  \hline
   $\eta_n$ & Regime & $\underset{n \rightarrow +\infty} \lim p_n$  \\
  \hline \hline
  $1$ & Free evolution & $1 - VT^2$   \\
  Constant $\in [0,1[$ & Zeno effect & 1  \\
  $1- \frac{\alpha}{n^{\beta}}$ with $\beta \in ]0,1[$ & Zeno effect & 1 \\
  $1- \frac{\alpha}{n^{\beta}}$ with $\beta>1$ & Free evolution &  $1 - VT^2$  \\
  $1- \frac{\alpha}{n}$ & Intermediate & $1 - \underbrace{2(\frac{1}{\alpha} + \frac{e^{-\alpha} -1}{\alpha^2})}_{\substack{\underset{\alpha \rightarrow +\infty} \longrightarrow 0 \text{ : Zeno effect} \\ \underset{\alpha \rightarrow 0} \longrightarrow 1 \text{ : free evolution}}} VT^2$  \\
   $1 - \alpha e^{-\beta n}$ & Free evolution & $1 - VT^2$   \\
  \hline
\end{tabular}
\\ \\

\section{Physical considerations concerning $\eta_n$} \label{physical}
\begin{enumerate}
\item As previously remarked, the constant case corresponds either to the absence of decoherence ($\eta = 1$) or to infinite decoherence ($\eta \in [0,1[$): these are not physically expected, except in some particular experimental setups (perfectly isolated systems for the first, experiments specifically designed to probe the Zeno effect for the second).

\item For now, we have not yet introduced any duration for the decoherence step, which was considered immediate. Let's at present assume that the time evolution can be divided into alternating steps dominated by either the free Hamiltonian, or by the interaction Hamiltonian. The time of interaction between the system and each environment $\mathcal{E}^i$, governed by $\hat{H}_{\mathcal{S}\mathcal{E}^i}$ of variance $\mathrm{Var}( \hat{H}_{\mathcal{S}\mathcal{E}^i}) \equiv V_{\text{int}}^ i \equiv V_{\text{int}}$ (constant strength of interaction), is still taken proportional to $\frac{T}{n}$, say equal to $\frac{cT}{n}$. This is a new assumption we make: that the time increments of both steps scale as $\frac{1}{n}$, and that the two phases can be considered on an equal footing, means that the two Hamiltonians are of relatively comparable strength. Then the quadratic approximation recalled in the introduction can be applied to the whole \{system + environment\}. Thus $\lvert \braket{ \mathcal{E}^i_\text{init} \vert \mathcal{E}^i_0} \rvert^2 = \lvert \braket{ 0 \mathcal{E}^i_\text{init} \vert 0 \mathcal{E}^i_0} \rvert^2 \simeq 1 -  V_{\text{int}} \left( \frac{cT}{n} \right)^2$. Since moreover $\braket{ \mathcal{E}^i_\text{init} \vert \mathcal{E}^i_0}$ is close to the real number 1 (infinitesimal decoherence), $\Re(\braket{\mathcal{E}^i_\text{init} \vert \mathcal{E}^i_0}) \simeq \lvert \braket{ \mathcal{E}^i_\text{init} \vert \mathcal{E}^i_0} \rvert \simeq  1 - \frac{1}{2}V_{\text{int}} \left( \frac{cT}{n} \right)^2$ is quadratic in time, and similarly for $\Re(\braket{\mathcal{E}^i_\text{init} \vert \mathcal{E}^i_1})$. This will also be the case for $\eta_n = \lvert \braket{\mathcal{E}^i_{0} \vert \mathcal{E}^i_{1}} \rvert$, because:

\begin{align*}
\sqrt{2-2\lvert \braket{\mathcal{E}^i_0 \vert \mathcal{E}^i_1} \rvert}  \simeq \sqrt{2-2\Re(\braket{\mathcal{E}^i_0 \vert \mathcal{E}^i_1 })} 
&= \lVert  \ket{\mathcal{E}^i_0 } - \ket{\mathcal{E}^i_1} \rVert \\
&\leq  \lVert  \ket{\mathcal{E}^i_\text{init}} - \ket{\mathcal{E}^i_0} \rVert +  \lVert \ket{\mathcal{E}^i_\text{init}} - \ket{\mathcal{E}^i_1} \rVert \\
&= \sqrt{2-2\Re(\braket{\mathcal{E}^i_\text{init} \vert \mathcal{E}^i_0})} + \sqrt{2-2\Re(\braket{\mathcal{E}^i_\text{init} \vert \mathcal{E}^i_1})} \\
&\simeq 2 \sqrt{V_{\text{int}}} \frac{cT}{n},
\end{align*}
hence $\eta_n = \lvert \braket{\mathcal{E}^i_{0} \vert \mathcal{E}^i_{1}} \rvert \gtrsim 1 - 2 V_{\text{int}} \left( \frac{cT}{n} \right)^2$ is also at least quadratic. Said differently, \textit{because quantum mechanical short time evolutions are always quadratic, and this is true also for the environment's evolution, infinitesimal steps of decoherence induced on a system by its surrounding environment are likely to be of the form $\eta_n = 1- \frac{\alpha}{n^{\beta}}$ with $\beta \gtrsim 2$. This could constitute a universal reason why the universe is not frozen by the quantum Zeno effect.} \\

An example of such an interaction is the following. Consider that the system is a qbit in the state $\frac{\ket{0} + \ket{1}}{\sqrt{2}}$, and the environment is a particle initially centered around $x=0$ with momentum $p_0$, that is $\ket{\Psi_{0, p_0}} = \frac{1}{\sqrt{\sqrt{\pi} \sigma}} e^{ip_0x} e^{-\frac{x^2}{2\sigma^2}} \in L^2(\mathbb{R})$. The system's state is recorded in the $(\ket{0} , \ket{1})$ basis due to the interaction $\hat{H}_{\mathcal{S}\mathcal{E}} = v \hat{\sigma_z} \otimes \hat{P}$ so that, after some time $\delta \propto \frac{1}{n}$:
\[ \left\{ \begin{array}{ll}
 \ket{0} \ket{\Psi_{0, p_0}} \longrightarrow \ket{0}\ket{\Psi_{v \delta, p_0}} \\ \\
\ket{1} \ket{\Psi_{0, p_0}} \longrightarrow \ket{1}\ket{\Psi_{-v\delta, p_0}} ,
 \end{array} \right. \] 
Here, 
\[ \eta_n = \lvert \braket{\Psi_{v \delta, p_0}  \vert \Psi_{-v\delta, p_0}} \rvert = \left\lvert \frac{e^{2ip_0v\delta}}{\sqrt{\pi} \sigma} \int_\mathbb{R} e^{-\frac{x^2+(v\delta)^2}{\sigma^2}} \mathrm{d}x \right\rvert = e^{-\frac{(v\delta)^2}{\sigma^2}} \simeq 1 - \frac{v^2}{\sigma^2} \delta^2, \]
so this interaction induces indeed a short time quadratic decoherence as long as the increment of time satisfies $\delta \ll \frac{v}{\sigma}$. \\

\item What if the assumption of comparable strengths of the Hamiltonians fails, for instance if the free evolution term is negligible compared to the coupling with the environment?  This amounts to taking $c$ or $V_{\text{int}}  \longrightarrow +\infty$, hence to lift the quadratic approximation for the interaction Hamiltonian. A possibility then is to consider that $\ket{\mathcal{E}^i_{0}(t)}$ and $\ket{\mathcal{E}^i_{1}(t)}$ follow two independent Brownian motions starting in $\ket{\mathcal{E}^i_\text{init}}$ on the sphere of all possible states in $\mathcal{H_{ \mathcal{E}^{\textit{i}}}}$ during the duration $\delta$ of the decoherence step. If the latter exceeds the typical time of diffusion on the sphere, we recover the case of a constant $\eta \in [0,1[$ (infinite decoherence, case n°1 above) with $\eta \sim \frac{1}{\sqrt{ \dim(\mathcal{H_{\mathcal{E}^{\textit{i}}}})}}$ as shown in \cite{soulas2024decoherence}. If it is shorter than the diffusion time (but still longer than the quadratic regime), $\ket{\mathcal{E}^i_{0}(\delta)}$ lies in the vicinity of $\ket{\mathcal{E}^i_\text{init}}$ on the sphere, which is approximately a ball. It is well known that the typical length of diffusion goes as $\lVert  \ket{\mathcal{E}^i_\text{init}} - \ket{\mathcal{E}^i_0(\delta)} \rVert \simeq D \sqrt{\delta}$, which implies $\lvert \braket{\mathcal{E}^i_\text{init} \vert \mathcal{E}^i_{0}(\delta)} \rvert \simeq \sqrt{1-(D \sqrt{\delta})^2} \simeq 1 - \frac{D^2}{2} \delta$. If $\delta$ is still taken $\propto \frac{1}{n}$, we are now in the intermediate regime studied above, with $\beta = 1$ and $\alpha \propto D^2$. This corresponds to situations where the system's evolution is slowed down because of its monitoring by the environment. In the limit of strong interaction, the diffusion constant $D$ will go to infinity and we recover the Zeno effect, whereas a weak interaction tends to the free evolution case. 
\end{enumerate}

\section{Comparison with Presilla \textit{et al.}'s continuous model} \label{comparison}

The situation explored by Presilla, Onofrio and Tambini in \cite[\S IV.]{presilla1996measurement} is very similar to ours — a two-level system undergoing some external monitoring in addition to its free evolution — but the major difference relies in the continuous nature of the model. Precisely, they solve the Lindblad's master equation governing the competition between the system's internal evolution in presence of a resonant field (producing Rabi oscillations between levels 1 and 2) and a continuous probing of the occupancy of level 1, whose intensity can modulated by a factor $\kappa(t)$. The equation, which can in this case be solved analytically, reads:

\[ \frac{\mathrm{d}\rho(t)}{\mathrm{d}t} = -\frac{i}{\hbar} \left[ \hat{H}(t) , \rho(t) \right] -\frac{1}{2} \kappa(t)  \left[ \hat{A} ,  \left[ \hat{A} , \rho(t) \right] \right], \]
where $\hat{A}$ is the monitored observable.

\begin{figure}[h]
\hfill
\subfigure[$\kappa=0$]{\includegraphics[scale=0.55]{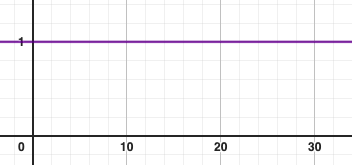}}
\hfill
\subfigure[$\kappa=0.2$]{\includegraphics[scale=0.55]{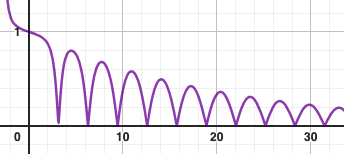}}
\hfill
\subfigure[$\kappa=1$]{\includegraphics[scale=0.55]{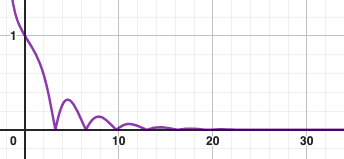}}
\hfill
\subfigure[$\kappa=5$]{\includegraphics[scale=0.55]{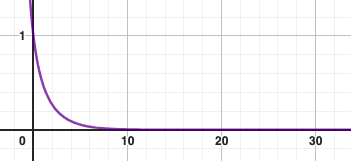}}
\hfill
\caption{Plot of $\eta(t)$ for different values of $\kappa$}
\label{fig2}
\end{figure}

It is interesting to check whether their findings are compatible with our model, in particular to determine the level of short time decoherence $\eta$ and the parameter $\beta$ that correspond to their situation and see if our conclusions still apply. To do so, recall that the density matrix of a system entangled with an environment in a state $\ket{\Psi} = c_1\ket{1}\ket{\mathcal{E}_1} + c_2\ket{2}\ket{\mathcal{E}_2}$ is given by $\rho = \begin{pmatrix}  \lvert c_1 \rvert^2 &  c_1 \bar{c_2} \braket{\mathcal{E}_2 \vert \mathcal{E}_1} \\  \bar{c_1} c_2 \braket{\mathcal{E}_1 \vert \mathcal{E}_2} &  \lvert c_2 \rvert^2 \end{pmatrix}$. Therefore, although the environment's evolution is not specified in Presilla \textit{et al.}'s model, one can still deduce $\lvert \braket{\mathcal{E}_1 \vert \mathcal{E}_2} \rvert$, and in particular its short-time behaviour corresponding to our $\eta$, by computing the quantity: $\eta(t) = \frac{\lvert \rho_{12}(t) \rvert}{\sqrt{ \rho_{11}(t) (1-  \rho_{11}(t)) }}$. Using the expressions (84) and (85) for $\rho_{11}(t)$ and $\rho_{12}(t)$ given in \cite{presilla1996measurement}, we can plot $\eta(t)$ for different values of the parameters. In Figure \ref{fig2}, we show $\eta(t)$ for the initial conditions $\rho_{11}(0)=1$, $\rho_{12}(0)=\rho_{21}(0)=0$, corresponding to a system starting in the pure state $\ket{1}$, and different values of $\kappa$ (we have set the Rabi frequency $\omega_R=1$, so that we actually plot in units of $\frac{\kappa}{\omega_R}$ and $\omega_R t$).

Crucially, we see that $\eta(t)$ has a non-vanishing first derivative in 0 (except in the extreme case $\kappa=0$, where of course no decoherence occurs). When plotting the latter as a function of $\kappa$, we observe that $\eta'(0)=-\frac{\kappa}{2}$. This means that the short-time decoherence is of the form $\eta(\delta) \simeq 1 - \frac{\kappa}{2} \delta$, which corresponds to the case $\beta = 1$ and $\alpha = \frac{\kappa}{2}$ in our model (recalling that $\delta = \frac{T}{n}$ and equating $T=1=\omega_R^{-1}$ the characteristic times of the two models).

Are the two models consistent? According to the table of Section \ref{analysis}, we expect an intermediate regime between Zeno freezing and free evolution, the system still evolving but slowed down. This is indeed what happens in Presilla \textit{et al.}'s model. Importantly, they conclude that ‘strong Zeno inhibition [$\kappa \rightarrow \infty$ limit] as well as full Rabi oscillations [$\kappa \rightarrow 0$ limit] are two trivial extreme regimes. However, [their model] tells us that another interesting and unexplored regime exists. It is the regime which occurs when the measurement coupling is comparable to the critical value. In this case a strong competition between stimulated transitions and measurement inhibition takes place’. 

We can even be more precise and compare quantitatively the slowing down of the system as a function of $\alpha = \frac{\kappa}{2}$. In our model, the quadratic evolution term $1-VT^2$ is replaced by $1 - 2(\frac{1}{\alpha} + \frac{e^{-\alpha} -1}{\alpha^2}) VT^2$, so that the time unit is effectively replaced by:
\[ T \rightsquigarrow T_\text{eff} = \frac{T}{\sqrt{2\left(\frac{1}{\alpha} + \frac{e^{-\alpha} -1}{\alpha^2} \right)}} = \frac{T}{2 \sqrt{ \frac{1}{\kappa} + 2\frac{e^{-\kappa/2} -1}{\kappa^2}}} \]
In the continuous model, the proper frequency $\omega_R$ of the system becomes under monitoring $\sqrt{\omega_R^2 - \frac{\kappa^2}{16}}$, so that the time unit is effectively replaced by:
\[ T \rightsquigarrow T_\text{eff} = \frac{T}{\sqrt{1 - \frac{\kappa^2}{16}}}. \]
Of course, these two correction factors are not equal, but they turn out to coincide quite well, at least in the small $\kappa$ case. For instance, on the whole range $\kappa \in [0, 2.5]$, they deviate by no more than 5\% to each other, as can be seen by plotting the function $F : x \mapsto \frac{2 \sqrt{ \frac{1}{x} + 2\frac{e^{-x/2} -1}{x^2}}} {\sqrt{1 - \frac{x^2}{16}}}$. For larger $\kappa$, though, the two models completely disagree at the quantitative level. This may be due to the fact that our model, by considering free evolution steps of finite time but immediate steps of decoherence, implicitly assumes some relatively weak coupling with the environment. In particular, it is unable to account for the critical transition that happens at $\kappa = 4$.

As a final comment in this section, it is interesting to remember that a continuous Lindblad's master equation corresponds to the intermediate regime $\beta=1$. This does not necessarily affects the argument given in Section \ref{physical} to justify that the most typical physical situation might be $\beta \gtrsim 2$. Indeed, although Lindblad's equation is somehow universal (being the most general equation governing an open quantum system interacting with a Markovian environment), picking a preferred observable $\hat{A}$ is not. In reality, the mutual monitoring between all the subsystems in the universe arise from complex interactions between a huge number of particles and it far from obvious to determine which observables are more recorded than the others (this discussion is related to the famous preferred-basis problem, see \cite{blog_riedel} for an updated bibliography).

\section{Discussion} \label{discussion}
We have presented a model designed to check whether quantum mechanics indeed predicts that the universe should evolve. To remain as universal as possible, no specific form of Hamiltonian was assumed. It allowed to determine the level of decoherence (induced by a surrounding environment) needed to freeze a two-level quantum system, arguably the kind of system the most prone to the Zeno effect. We have found that if, during a time interval $\frac{T}{n}$, the environment distinguishes between the two states according to $\lvert \braket{\mathcal{E}^i_{0} \vert \mathcal{E}^i_{1}} \rvert  \simeq 1- \frac{\alpha}{n^{\beta}}$ with $\beta>1$, then free evolution wins over decoherence and the system is not frozen. In the most generic case, because quantum mechanical short time evolutions are always quadratic (and this is true for the system as well as for the pair \{system + environment\}), we find $\beta \gtrsim 2$, hence the universe is indeed not frozen. We have finally made a quantitative comparison with the continuous master equation model by Presilla, Onofrio and Tambini \cite{presilla1996measurement}. The links between the two models are non-trivial but we have found a good agreement at least in the low coupling regime.

The main weaknesses of the model, leading to possible improvements, are the following.
\begin{itemize}
\item Is the discrete setup legitimate? A succession of infinitesimal steps is not necessarily the same as a joint continuous evolution.
\item What happens if the coupling with the environment is not supposed roughly constant anymore? Mathematically, this means that the $\eta^i$ are not equal, and the infinitesimal decoherence rate (\textit{i.e.} the flow of information) at time $t$ could be modelled in the limit $n \longrightarrow +\infty$ as a continuous quantity $1-\mathrm{d}\eta(t)$. It is natural to ask for the set of such functions which entail a Zeno freezing. Besides, the durations of the steps could also be non-constant (like following a Poisson process, as done in \cite{kulkarni2023first}).
\item Assuming the environments $\mathcal{E}^i$ distinct and non-entangled is a very unphysical assumption. In some cases, previous entanglement among the environments can dramatically change the efficiency of decoherence. As an example, take an environment composed of two qbits called $\mathcal{E}^1$ and $\mathcal{E}^2$ initially maximally entangled; the system interacts with $\mathcal{E}^1$ then with $\mathcal{E}^2$ \textit{via} a C-NOT gate:

\begin{align*}
\underbrace{\frac{1}{\sqrt{2}} (\ket{0} + \ket{1}) }_{\mathcal{S}} \otimes \underbrace{ \frac{1}{\sqrt{2}} (\ket{0 0} + \ket{1 1}) }_{\mathcal{E}^1+\mathcal{E}^2}
&\quad \underset{C-NOT_{\mathcal{S}\mathcal{E}^1}} \longrightarrow \quad \underbrace{\frac{1}{2} (\ket{0 0 0} + \ket{ 0 1 1} + \ket{1 1 0} +\ket{1 0 1} )}_{\rho_{\mathcal{S}} =  \begin{pmatrix} \frac{1}{2} & 0 \\ 0 & \frac{1}{2} \end{pmatrix}\text{ : $\mathcal{S}$ is perfectly decohered}}  \\
&\quad \underset{C-NOT_{\mathcal{S}\mathcal{E}^2}} \longrightarrow \quad \frac{1}{2} (\ket{0 0 0} + \ket{ 0 1 1} + \ket{1 1 1} +\ket{1 0 0}) \\
&\hskip2cm = \underbrace{ \frac{1}{\sqrt{2}} (\ket{0} + \ket{1}) \otimes  \frac{1}{\sqrt{2}} (\ket{0 0} + \ket{1 1}).}_{\rho_{\mathcal{S}} =  \begin{pmatrix} \frac{1}{2} & \frac{1}{2} \\ \frac{1}{2} & \frac{1}{2} \end{pmatrix}\text{ : coherence has revived}} 
\end{align*} 
\end{itemize}

\newpage
\bibliographystyle{siam}
\bibliography{Biblio_Zeno}
\end{document}